\documentclass[12pt]{article}
\usepackage{a4wide}
\usepackage{amsfonts}

\begin{document}

\begin{flushright}
{}FTUV 06/2212 \quad IFIC 06-46
\end{flushright}
\vskip0.5cm

\begin{center}
\begin{large}
{\bf On D=11 supertwistors,  superparticle quantization and a
hidden $SO(16)$ symmetry of supergravity}\footnote{Invited
contribution at the XXII Max Born Symposium {\it Quantum, Super
and Twistors}, September 27-29, 2006 Wroclaw (Poland)}.
\end{large}

 \vskip1cm

{\bf Igor A. Bandos${}^{a,b}$, Jos\'e A. de Azc\'arraga${}^a$ and
Dmtri Sorokin${}^{a,c}$}
 \vskip.5cm
 {\it ${}^a$ Dept.Theoretical Physics, Valencia University, \\and
IFIC (CSIC-UVEG), Spain\footnote{\small bandos@ific.uv.es,
j.a.de.azcarraga@ific.uv.es,
dmitri.sorokin@ific.uv.es}}\\
{\it ${}^b$ ITP  KIPT Kharkov, Ukraine} \\
{\it  ${}^c$ INFN and Dept. of Physics, Padova Univ., Italy} \\

\end{center}

\begin{abstract}
We consider a covariant quantization of the $D$=11 massless
superparticle in the supertwistor framework.  $D$=11 supertwistors
are highly constrained, but the interpretation of their bosonic
components as Lorentz harmonic variables and their momenta permits
to develop a classical and quantum mechanics without much
difficulties. A simple, heuristic `twistor' quantization of the
superparticle leads to the linearized $D$=11 supergravity
multiplet. In the process, we observe hints of a hidden $SO(16)$
symmetry of $D=11$ supergravity.
\end{abstract}

\section{Introduction}

Let us begin by thanking the organizers of the XXII Max Born
symposium, in honour of our friend Jerzy Lukierski, for their
invitation. The title of this contribution is in one-to-one
correspondence with the topics of the symposium: {\it Quantum,
Super and Twistors}. We present here a {\it twistor quantization}
of the $D$=11 massless {\it super}particle. But this coincidence
is not our only motivation. Recently, the covariant description of
the quantum superstring in the `pure spinor' approach of Berkovits
(applied also to superparticles and the $D$=11 supermembrane) has
led to the first results in superstring covariant loop
calculations (see \cite{NBloops} and refs. therein). In spite of
the present progress in understanding the relation \cite{pure-GS}
of the pure spinor superstring \cite{NBloops} with the original
Green-Schwarz formulation, as well as \cite{Dima+Mario+02} with
the superembedding approach \cite{bpstv,Dima}, a further study of
the origin and geometrical meaning of the pure spinor formalism as
well as of its possible modifications (see {\it e.g.}
\cite{nonmNB}) seems appropriate.

In this respect the Lorentz harmonics approach
\cite{Sok,Lharm,Ghsds,BZ,BZstrH,GHT93} may be of interest, since
some progress toward a covariant superstring quantization had
already been made in the late eighties \cite{Lharm} in that
context. Although nothing like the recent breakthrough in loop
calculations \cite{NBloops} has been achieved in such a framework,
its connection with the superembedding approach \cite{bpstv,Dima},
its geometrical meaning \cite{Sok,Ghsds,BZ,BZstrH}, and its
relation with twistors \cite{BZ,BZstrH} suggests applying it to
the covariant superstring quantization (in the pragmatic spirit of
the pure spinor approach of \cite{NBloops}, rather than attempting
to develop a full-fledged hamiltonian approach as in
\cite{BZstrH,JdA00b}). To this aim, a natural first step is the
covariant quantization of the massless superparticle. We sketch it
here for the $D$=11 case, which leads to the $D$=11 supergravity
multiplet. Recently, there has been an extensive search for hidden
symmetries of M-theory; our analysis gives further evidence for a
hidden $SO(16)$ symmetry of $D$=11 supergravity \cite{Nicolai87}.

\section{D=11 massless superparticle action in the spinor
moving frame (Lorentz harmonics) formulation}

The action of the $D=11$ massless superparticle
\cite{B+T=D-branes} can be presented in the following equivalent
forms \cite{BL98'}
\begin{eqnarray}
\label{11DSSP} S&:= \int d\tau L =   \int_{W^1} {1\over
2}\rho^{++}\, u_{m}^{--} \, \Pi^m = \int_{W^1} {1\over
32}\rho^{++}\, v_{\alpha q}^{\; -} v_{\beta q}^{\; -} \, \Pi^m
\tilde{\Gamma}_m^{\alpha\beta} =
\\ \label{11DSSP+}
& =\int_{W^1}\rho^{++}\, v_{\alpha q}^{\; -} v_{\beta q}^{\; -} \,
\Pi^{\alpha\beta} \qquad (\alpha=1,\dots,32\;,\; q=1,\dots,16) \;,
  \end{eqnarray}
where $\alpha$ is the $D$=11 spinor index, $q$ is the $SO(9)$
spinor index and $-,--,++$ denote $SO(1,1)$ representations (or
scaling dimensions --1,--2,+2 respectively). In (\ref{11DSSP}),
the action is given in terms of the invariant one-form $\Pi^m$ on
{\it standard} superspace $\Sigma^{(11|32)}$, $\Pi^m := dx^m -
id\theta\Gamma^m\theta$, and in (\ref{11DSSP+}) in terms of the
Cartan form $\Pi^{\alpha\beta}= \Pi^{\beta\alpha}$ on
 the {\it maximally enlarged, tensorial superspace} $\Sigma^{(528|32)}$,
\begin{eqnarray}
\label{PiG}
 \Pi^{\alpha\beta}&=&  dX^{\alpha\beta} - i d\theta^{(\alpha}\, \theta^{\beta)}= \qquad
 \\ &=& {1/32} \Pi^m \tilde{\Gamma}_m^{\alpha\beta}
 - {1/64} i \Pi^{mn} \tilde{\Gamma}_{mn}^{\alpha\beta} + {1/(32\cdot\, 5!)}
 \Pi^{m_1\ldots m_5}
 \tilde{\Gamma}_{m_1\ldots m_5}{}^{\alpha\beta} \; , \quad \nonumber
\end{eqnarray}
which includes 517 extra bosonic tensorial coordinates (see
\cite{van2, BL98,BL98',JdA00} and refs. therein) in addition to
the eleven spacetime coordinates $x^m =
\Pi^{\alpha\beta}\Gamma^m_{\alpha\beta}$ and the 32 fermionic ones
of $\Sigma^{(11|32)}$. Tensorial  $\Sigma^{({n(n+1)\over 2}|n)}$
superspaces were used to describe $D$=4,6 and 10 conformal higher
spin fields through the quantization of the tensorial
superparticle \cite{BL98} (see \cite{BLS99+BBdAST04} and refs.
therein and \cite{JdA00b} in the BPS preons \cite{BPS01} context).

The equivalence of the above two seemingly different forms of the
superparticle action occurs due to the following constraints that
are imposed on the spinor variables (see \cite{BZ,BZstrH} and
\cite{BL98'} for $D$=11),
\begin{eqnarray}\label{vv=uG}
\cases{2 v_\alpha{}_{q}^{-} v_\beta{}_{q}^{-} =
 u_m^{--}{\Gamma}^m_{\alpha\beta}\;\;(a)\;
 , \quad \cr  v_{q}^{-}\tilde{\Gamma}_m v_{p}^{-} = 2\, \delta_{qp} \; u_m^{--}  \; (b)\;, }
\;   v_\alpha{}_{q}^{-}C^{\alpha\beta}v_\beta{}_{q}^{-}=0\;\;
(c)\;,
   \quad u_m^{--}u^{m --}=0 \; (d)\;\; . \qquad
\end{eqnarray}
The first constraint (\ref{vv=uG}a) eliminates from the action
(\ref{11DSSP+}) the rank two and rank five tensorial Cartan forms
contributions, $\Pi^{mn}$ and $\Pi^{m_1\cdots m_5}$ in
(\ref{PiG}).

   Although, in principle, one can study the superparticle
dynamical system using just the constraints (\ref{vv=uG}) (see
\cite{gs92}), it is more convenient to look at the lightlike
vector $u_m^{--}$ (Eq. \ref{vv=uG}d) as an element of a {\it
vector} Lorentz moving frame and to treat the set of the sixteen
$SO(1,10)$ spinors $v_\alpha{}_{q}^{-}$ as part of the associated
{\it spinor} moving frame. These variables are called,
respectively, vector and spinor Lorentz harmonics\footnote{Note
that, in contrast to $D$=3,4,6,10 spacetimes, in which the vector
$p^m=\lambda\Gamma^m\lambda$ constructed using a single commuting
Majorana or (symplectic) Majorana-Weyl spinor $\lambda_\alpha$ is
automatically lightlike
 $[\,p^m p_m=(\lambda\Gamma^m\lambda)(\lambda\Gamma_m\lambda)=0\,]$
and thus can be identified with the momentum of a massless
particle, $(\lambda\Gamma^m\lambda)^2 \not= 0$ in $D=11$ for a
generic commuting Majorana spinor. Thus, to develop a twistor-like
description of a massless $D=11$ superparticle one has to
introduce {\it constrained} spinor variables; the Lorentz
harmonics $v^{-}_{\alpha q}$ are one of the possible choices.
Another one is given by the `pure' spinors of \cite{NBloops},
which are complex spinors $\Lambda_{{\alpha}} = w^1_{{\alpha}}+
iw^2_{{\alpha}}$ obeying $\Lambda \Gamma_{m}\Lambda =0$.}.

 {\it Vector} harmonic variables \cite{Sok} are defined as
 elements of the Lorentz $11\times 11$ matrix $[\,m=(0,1,\dots,9,\#)\; ;
 \;(a)=(--,\,++,\,I)\; ;\; I=(1,\dots,9)\,]$,
\begin{eqnarray}
\label{harmUin} U_m^{(a)}= (u_m^{--}, u_m^{++}, u_m^{I})\;  \in \;
SO(1,10) \quad ,
\end{eqnarray}
where $u^{\pm\pm}_m=u^0_m \pm u^{\#}_m$. The fact that $U\in
SO(1,10)$ implies the constraints
\begin{eqnarray}\label{harmUdef}
U^T\eta U = \eta  \quad \Leftrightarrow \cases{ u_m^{--}u^{m--}=0
\; , \quad u_m^{++}u^{m++}=0 \; , \quad u_m^{\pm\pm}u^{m\, i}=0 \;
, \cr u_m^{--}u^{m++}=2 \; ,  \qquad u_m^{i}u^{m\, j}=-
\delta^{ij} }
\end{eqnarray}
and/or equivalently,  $\delta_m^n= {1\over 2}u_m^{++}u^{n--} +
{1\over 2}u_m^{--}u^{n++} - u_m^{i}u^{n i}$ ($U\eta U^T=\eta$).

  Similarly, the {\it spinor}  harmonic \cite{Ghsds} or
spinor moving frame \cite{BZ} variables $v^{\pm}_\alpha{}_q$ are
the elements of the $Spin(1,10)$ $32\times32$ matrix
\begin{eqnarray}
\label{harmVin} V_\alpha^{(\beta)}= (v_\alpha{}_q^{-}\; ,
v_\alpha{}_{q}^{+})\; \in \; Spin(1,10)\quad (\alpha=1,\dots 32\;
, \; q=1,\dots,16) \; .
\end{eqnarray}
They are `square roots' of the associated vector harmonics in the
sense that
\begin{eqnarray}
\label{harmVdef} V \Gamma^{(a)} V^T = \Gamma^m u_m ^{(a)} \; ,
\qquad V^T \tilde{\Gamma}_m V = u_m^{(a)} \tilde{\Gamma}_{(a)}\; ,
\end{eqnarray}
which express the $Spin(1,10)$ invariance of the Dirac matrices.

Equation in (\ref{vv=uG}a) is just the $(a)=(--)\equiv (0)-(\# )$
component of the first equation in (\ref{harmVdef}) in the Dirac
matrices realization in which $\Gamma^0$ and $\Gamma^{\# }$ are
diagonal; the nine remaining $\Gamma^I$ are off-diagonal. Eq.
(\ref{vv=uG}b) comes from the upper diagonal block in the second
equation in Eq. (\ref{harmVdef}); the lightlike character of
$u_\mu^{--}\equiv u_\mu^{0}- u_\mu^{\# }$ (see (\ref{harmUdef}))
follows from the orthogonality and normalization of the timelike
$u_m^{0}$ and spacelike $u_m^{\# }$ vectors. To complete the set
of constraints defining the spinorial harmonics, we have to add
the conditions expressing the invariance of the charge conjugation
matrix $C$,
\begin{eqnarray}
\label{harmVdefC}
 VCV^T=C \quad, \quad V^TC^{-1}V=C^{-1}\; ,
\end{eqnarray}
which give rise to the constraint (\ref{vv=uG}c).

In a theory with a local $SO(1,1)\otimes SO(9)$ symmetry
containing only one of the two sets of $16$ constrained spinors
(\ref{harmVin}), say the spinors $v_{\alpha p}^{\;-}\,$, these can
be treated as homogeneous coordinates of the $SO(1,10)$ coset
giving the celestial sphere $S^9$; specifically (see \cite{Ghsds,
GHT93})
\begin{eqnarray}
\label{v-inS11} {} \{v_{\alpha q}^{\;-}\} = {Spin(1,10) \over
[SO(1,1)\otimes Spin(9)] \, \subset \!\!\!\!\!\!\times
{\mathbb{K}_9} } = \mathbb{S}^{9}  \quad ,
\end{eqnarray}
where $\mathbb{K}_9$ is the abelian subgroup of $SO(1,10)$ defined
by
 $\delta v_{\alpha q}^{\; -}=0\;,\; \delta v_{\alpha q}^{\; +}=
 k^{++}_i \gamma^i{}_{qp}\,v_{\alpha p}^{\; -}$. Our
superparticle model is of such a type.

\section{Supertwistor formulation of the
D=11 superparticle action and a first appearance of SO(16)}

Using the Leibnitz rule ($dx \, v v = d(xv) v - xdv\,v$ etc.) the
superparticle Lagrangian, $d\tau L$ in (\ref{11DSSP}), can be
written as the  twistorial Liouville form,
\begin{eqnarray}\label{S=rhoTw}
S &=&  \int_{W^1} (\lambda_{\alpha q}\,  d {\mu}^{\alpha}_{q}-
d\lambda_{\alpha q}\;  {\mu}^{\alpha}_{q} - i d\eta_{q}\,\eta_{q})
\; ,
\end{eqnarray}
where the sixteen 32-component spinors $\lambda_{\alpha q}$ are
taken to be proportional to the spinor harmonics $v_{\alpha
q}^{\;-}$,
\begin{eqnarray}
\label{Tw=H} \lambda_{\alpha q}:= \sqrt{\rho^{++}} v_{\alpha
q}^{\; -}\; . \qquad
\end{eqnarray}
Hence, they obey the constraints (see (\ref{vv=uG});
 $\Gamma^m\equiv \Gamma^m{}_{\alpha\beta}\,,\,
 {\tilde \Gamma}^m\equiv   {\tilde \Gamma}^m{\,}^{\alpha\beta}$)
\begin{eqnarray}\label{ll=pG}
2 \lambda_\alpha{}_{q} \lambda_\beta{}_{q} = p_m
{\Gamma}^m_{\alpha\beta}\; , \qquad \lambda_{q} \tilde{\Gamma}_m
\lambda_{p} = \delta_{qp} \; p_m  \; ,  \qquad C^{\alpha\beta}
\lambda_\alpha{}_{q}
   \lambda_\beta{}_{p}=0\; ,
\end{eqnarray}
where the particle momentum vector $p_m=\rho^{++}u_m^{--}$ is
lightlike due to (\ref{harmUdef}). On account of $\rho^{++}$ in
(\ref{Tw=H}), the $\{\lambda_\alpha{}_{q}\}$ parametrize the
$\mathbb{R}_+ \times \mathbb{S}^{9}$ manifold ({\it cf.}
(\ref{v-inS11})).
The variables $ {\mu}^{\alpha}_{q}$, $\eta_{q}$ in (\ref{S=rhoTw})
are related to the superspace coordinates by the following
generalization of the Penrose incidence relation,
\begin{eqnarray}\label{Tw=}
{\mu}^{\alpha}_{q}:= X^{\alpha\beta} \lambda_{\beta q} - {i\over
2} \theta^\alpha \, \theta^\beta \lambda_{\beta q}  \; , \qquad
\eta_{q}:= \theta^\beta \lambda_{\beta q} \; .
\end{eqnarray}
  Together with $\lambda_{\alpha q}\,$, the variables
${\mu}^{\alpha}_{q}$ and ${\eta}_{q}$ define a set of sixteen
constrained $OSp(1|64)$ supertwistors,
\begin{eqnarray}
\label{Y-q=}
 \Upsilon_{\Sigma \,q} := ( \lambda_{\alpha q}\; , \;
{\mu}^{\alpha}_{q}\; , \;  {\eta}_{q})\quad , \quad  \matrix{
q=1,\dots,16 \; , \; \alpha=1,\dots,32\;}  .
\end{eqnarray}
In terms of them the action (\ref{S=rhoTw}) reads
\begin{eqnarray}\label{stwaction}
S=\int_{W^1}  d{\Upsilon}_{\Sigma\,
q}\Omega^{\Sigma\Pi}\Upsilon_{\Pi\,q} \quad , \quad
\Omega^{\Sigma\Pi} = \left(\matrix{ 0 & \delta^\alpha{}_\beta & 0
\cr -\delta_\alpha{}^\beta & 0 & 0 \cr 0 & 0 & i } \right) \quad ,
\end{eqnarray}
where $\Omega^{\Sigma\Pi}$ is the orthosymplectic $OSp(1|64)$
metric.

 The supertwistors in this action are very constrained.
First, because the basic spinors $\lambda_{\alpha q}$ obey the
constraints in Eq. (\ref{ll=pG}) by virtue of Eqs. (\ref{Tw=H} ,
\ref{vv=uG}), and, secondly, since Eq. (\ref{Tw=}) provides the
general solution of (and, hence, can be replaced by) the following
set of constraints
\begin{eqnarray}
\label{SO(9)=} & {\mathbb{J}}_{pq}:=
 {\Upsilon}_{\Sigma\, p}\Omega^{\Sigma \Pi} {\Upsilon}_{\Pi\,q}
:=
 2 \lambda_{\alpha [p}\mu^{
\alpha}_{q]} + i {\eta}_{[p}{\eta}_{q]} =0 \, . \qquad
\end{eqnarray}

Notice that, if $X^{\alpha\beta}$ is restricted to include only
the spacetime coordinates in $\Sigma^{(11|32)}$,
$X^{\alpha\beta}\propto x^m \tilde{\Gamma}_m^{\alpha\beta}$ (see
(\ref{11DSSP})), one more set of constraints is found
\begin{eqnarray}
\label{moreII=} K_{pq}= K_{qp}:= \lambda_{\alpha (p}\;
{\mu}^{\alpha}_{q)} - {1\over 16}\delta_{pq}\, \lambda_{\alpha
p'}\, {\mu}^{\alpha}_{p'}= 0 \; .
\end{eqnarray}
When one starts from the equivalent form of the action
(\ref{11DSSP+}) with general $X^{\alpha\beta}=1/32 x^m
\tilde{\Gamma}_m^{\alpha\beta} + y^{mn}
\tilde{\Gamma}_{mn}^{\alpha\beta}+ y^{[5]}
\tilde{\Gamma}_{[5]}^{\alpha\beta}$, Eq. (\ref{moreII=}) appears
as a (partial set of) gauge fixing conditions for the gauge
symmetry $\delta \mu^{\alpha}_{q}= b^{mn}
(\tilde{\Gamma}_{mn}\lambda_q)^\alpha + b^{[5]}
(\tilde{\Gamma}_{[5]}\lambda_q)^\alpha$ of the action
(\ref{S=rhoTw0}). This gauge symmetry follows from the
consequences $v^-_q\Gamma^{[2]}v^-_q=0$,
$v^-_q\Gamma^{[5]}v^-_q=0$, of the constraint (\ref{vv=uG}a), and
may gauge away the contributions of the extra coordinates
$y^{mn}=y^{[2]}$ and $y^{mnpqr}=y^{[5]}$ in $X^{\alpha\beta}$ in
Eq. (\ref{Tw=}).

The above discussion indicates the second class character of the
constraint (\ref{moreII=}). In contrast, Eq. (\ref{SO(9)=}) can be
treated as a gauge symmetry generator. Namely, with Eqs.
(\ref{Tw=H}, \ref{v-inS11}), they are generators of the
$2^{[9/2]}=36$ parametric $Spin(9)$ symmetry. Actually, already at
this stage {\it it is tempting to treat} the ${\mathbb{J}}_{pq}$
in (\ref{SO(9)=}) {\it as $SO(16)$ generators}. For this to be the
case one should use
\begin{eqnarray}\label{Tw=HS}
\lambda_{\alpha q}:= \sqrt{\rho^{++}}(\tau) v_{\alpha p}^{\;
-}(\tau) S_{pq}(\tau)\; , \qquad S_{pq'}S_{qq'}=\delta_{pq} \;
,\quad S\in SO(16)\;,
\end{eqnarray}
instead of  (\ref{Tw=H}) as the definition of $\lambda_{\alpha
q}$. This is indeed possible since, if one substitutes $v_{\alpha
p}^{\; -}(\tau) S_{pq}(\tau)$ with $S_{pq}\in SO(16)$ for
$v_{\alpha q}^{\; -}$ in (\ref{11DSSP}), the two $SO(16)$ factors
cancel ($SS^T=1$) and the action remains the same.

We will return to the question of this $SO(16)$ symmetry, already
suggested by the existence of an $SO(16)$ covariant formulation of
$D$=11 supergravity \cite{Nicolai87}, after discussing the
twistorial quantization of the dynamical system of
(\ref{S=rhoTw}).

\section{Supertwistor covariant quantization in D=11}

The dynamical variables in the action (\ref{S=rhoTw}) or
(\ref{stwaction}) are highly constrained, promising to make a full
Hamiltonian analysis of the constraints rather involved (see
\cite{BZstrH, JdA00b}). However, the group theoretical meaning of
these constrained variables allows us to present a simple
alternative, the {\it twistor covariant quantization}.

{\it 3.1.Twistor quantization of the $D$=11 massless bosonic
particle}.

In the purely bosonic case $\eta_q=0$ and the lagrangian in
(\ref{S=rhoTw}) reduces to
\begin{eqnarray}
\label{S=rhoTw0} S_b &=&  \int_{W^1} (\lambda_{\alpha q}\,  d
{\mu}^{\alpha}_{q}-
 d\lambda_{\alpha q}\;  {\mu}^{\alpha}_{q} ) \; .
\end{eqnarray}
As we saw, the constrained (Eqs. (\ref{ll=pG})) spinors
$\lambda_{\alpha q}$
 parametrize the celestial $\mathbb{S}^9$
sphere times $\mathbb{R}_+$. Due to the first two equations in
(\ref{ll=pG}), this manifold can be identified with that of the
positive zero mass shell momenta,
\begin{eqnarray}
\label{p-inS11} & \Omega_0^+(p) := {} \{p_m \, : \;
p^2=0\;,\;p_0>0 \} = \mathbb{R}_+\times \mathbb{S}^{9} \; . \qquad
\end{eqnarray}
The variables ${\mu}^{\alpha}_{q}$  in (\ref{S=rhoTw}) are clearly
identified as those canonically conjugated to
 $\lambda_{\alpha q}$, and should be constrained as the $\lambda_{\alpha q}$
are. Thus the seemingly simple dynamics in (\ref{p-inS11}) is
actually quite complicated. However, if all gauges were fixed and
all the second class constraints were solved to have all
$\lambda_{\alpha q}$ expressed through the ten coordinates of
$\mathbb{R}_+\; \times \mathbb{S}^{9}$, then
 the set $\mu^{\alpha}_q$ would just contain the corresponding
ten independent conjugate variables.

Thus, the quantization of the system described by {\it the
twistorial Liouville action} (\ref{S=rhoTw0}) leads, in the
simplest case, to scalar wavefunctions $\Phi(\lambda)$ with
arguments on $\mathbb{R}_+\otimes \mathbb{S}^9$. By
(\ref{p-inS11}), this manifold can be identified with the $D$=11
lightlike momenta `cone' $\Omega^+_0(p)$. Hence, the wavefunctions
$\Phi(\mathbb{R}_+\times \mathbb{S}^9)=\Phi (p_m ; \, p^2=0,\,
p_0>0)$ describe positive energy solutions of the $D$=11 massless
Klein-Gordon equation.

The scalar wavefunction $\Phi$ is invariant under the $SO(9)$
symmetry, but this is not the only possibility. The $SO(9)$ gauge
symmetry which defines the basic variables as `homogeneous'
coordinates of $\mathbb{R}_+ \times\mathbb{S}^{9}$, also allows
for wavefunctions that transform non trivially under
 $SO(9)$ as {\it e.g.}, $SO(9)$ spinors $\Psi_q(p_m ; p^2=0)=
 <p_m|q>\;,$
vectors $\Phi_I(p_m ; p^2=0)$= $<p_m|I>$, etc. In the $D$=4 case
these would be associated with the different choices of the
constants ({\it e.g.} the helicity for the analogous to
(\ref{SO(9)=})) in the quantum constraints with non-commuting
operators; it would be interesting to understand this in the
higher dimensional $D$=10, 11 cases.

{\it 3.2. Twistor quantization of the D=11 massless
superparticle}.

In the supersymmetric case the action (\ref{S=rhoTw}) is provided
by the sum of the bosonic action (\ref{S=rhoTw0}) and the free
fermionic one,
\begin{eqnarray}
\label{S=S0+Sf} & S =  S_b - \int_{W^1} i d\eta_{q}\,
 \eta_{q} \; .
\end{eqnarray}
The fermions are decoupled from the bosonic $\mathbb{R}_+ \times
\mathbb{S}^9$ part. The expression of the fermion canonical
momentum is a second class constraint that identifies $\eta_q$
with its own momentum. Thus one has to use Dirac starred brackets
and their associated quantum anticommutators,
\begin{eqnarray}
\label{epeqD=}
 & {}\{ \eta_{q}\, , \,
 \eta_{p} \}^* = {i\over 2}  \delta_{qp}\quad \rightarrow \quad
  \{ \hat{\eta}_{q}\, , \,
 \hat{\eta}_{p} \} =  {1\over 2}  \delta_{qp} \quad,
\end{eqnarray}
which imply that {\it the $\hat{\eta}_{q}$ generate the
$SO(16)$-invariant Clifford algebra.}

This fact seems to reflect the hidden $SO(16)$ symmetry of $D$=11
supergravity \cite{Nicolai87}.
 The natural representation of $\hat{\eta}_{q}$ is given by the
 $256\times256$ $SO(16)$ gamma matrices,
$\hat{\eta}_{q}=({\bf\Gamma}_q)^{\cal A}{}_{\cal B}$, which act on
the Majorana spinors of $SO(16)$. This representation decomposes
under $SO(16)$ into the sum ${\bf 128}\oplus {\bf 128}$ of two
Majorana-Weyl spinors and under $SO(9)$ as ${\bf 256}= {\bf
128}\oplus{\bf 44}\oplus{\bf 84}$, corresponding to an $SO(9)$
spin-tensor $|qI>$, a symmetric traceless second rank tensor $|IJ>
\, \equiv |(IJ)> $ and an antisymmetric third rank tensor $|IJK>
\,\equiv |[IJK]> $. Thus one may introduce  a natural Grassmann
grading (with $|qI>$ fermionic) and consider the following
representation \cite{Green+99} of the $SO(16)$-invariant Clifford
algebra (\ref{epeqD=}):
\begin{eqnarray}
\label{repr-16}
&& {2}\hat{\eta}_{q} |IJ> = \Gamma^I_{qp} |pJ>+
\Gamma^J_{qp} |pI> \; , \qquad
\nonumber \\
&& {2}\hat{\eta}_{q} |pI> = {1\over 2} \Gamma^I_{qp} |IJ>+ {1\over
3\cdot 4!} \left(\Gamma^{IJ_1J_2J_3}_{qp} - 6 \delta^{I[J}
\Gamma^{J_2J_3]}_{qp}\right) |J_1J_2J_3> \; ,
\qquad \nonumber \\
&&  {2}\hat{\eta}_{q} |IJK> =  \Gamma^{IJ}_{qp} |pK>+
\Gamma^{KI}_{qp} |pJ>+ \Gamma^{JK}_{qp} |pI> \; . \qquad
\end{eqnarray}

This representation allows us to conclude that the {\it
quantization of the $D=11$ massless superparticle model
(\ref{11DSSP}) reproduces the D=11 supergravity multiplet} (see
\cite{Green+99} for a light-cone gauge analysis and \cite{NBloops}
for the pure spinor quantization). Such a multiplet is described
by the following set of wavefunctions on $\Omega_0^+(p)$ (Eq.
(\ref{p-inS11})) corresponding to (\ref{repr-16}),
\begin{eqnarray}\label{PsiIq-}
& \cases{ \Psi_{I\,q}(p)= <p^m_{p^2=0} | qI> \quad , \quad [p_m
p^m=0] \;  \cr h_{(IJ)}(p)= <p^m_{p^2=0} |IJ> \; , \quad
A_{IJK}(p)= <p^m_{p^2=0} |IJK> \; , } \quad
\end{eqnarray}
which give the general solution of the linearized $D$=11
supergravity equations. Each of these wavefuntions provides an
irreducible non-trivial representation of the $SO(9)$ part of the
`little group' of a $D$=11 lightlike momentum. Schematically, the
manifestly covariant solution of the field equations in momentum
space, in terms of the transverse ($u^I_m p^m=0$) vector harmonic
and the $\lambda_q$ spinor variables (Eqs. (\ref{harmUin}) and
(\ref{Tw=H}) respectively), becomes
\begin{eqnarray}
\label{Psi-muA} & \cases{\Psi_{m\,\alpha}(p) = \Psi_{I\,
q}(p)u^I_m\; \lambda_{\alpha q}\; , \quad [\lambda_{\alpha q}
\lambda_{\beta q}= {1\over 2} \Gamma^m_{\alpha\beta}p_m \;, \;
p^2=0 \; , \; p^m u^I_m=0] \,\cr
 h_{mn}(p)= u^I_m u^J_n h_{(IJ)}(p)\;,
 \qquad A_{m n p}(p)= u^I_m u^J_n u^K_p A_{IJK}(p) \; .  }
\end{eqnarray}
Similar solutions were discussed in \cite{GHT93}, which is devoted
to a twistor transform of the linearized field equations in
various dimensions.

\section{Conclusions and outlook}

We have outlined here a covariant quantization of the $D$=11
massless superparticle in the Lorentz harmonics formalism. This
quantization is twistor-like in the sense that we have used a
supertwistor form for the massless superparticle action
\cite{BL98'}, itself a $D$=11 counterpart of the Ferber-Shirafuji
one \cite{F78S83}. In contrast with the $D$=4 case, the $D$=11
supertwistors are very constrained variables (see \cite{BZ,BZstrH}
and the more recent \cite{Itzhak+MP05+IB+JdA+C06} for further
discussion), but their group-theoretical meaning allows us to
handle this problem. Our analysis indicates, in particular, a
possible origin (see
Eqs.(\ref{SO(9)=},\ref{Tw=HS},\ref{epeqD=},\ref{repr-16}))
 for the hidden $SO(16)$ symmetry of $D$=11 supergravity
\cite{Nicolai87}.

An interesting direction for further development is to look for a
Lorentz harmonics version of the BRST quantization. It would also
be interesting to see whether and how the $D$=11 supergravity
supermultiplet appears when one uses other representations for the
operator algebra (\ref{epeqD=}) as {\it e.g.}, when wavefunctions
are described by a Clifford superfield (see \cite{Dima88}), $
\mathbb{W}=\mathbb{W}(p_m , \hat{\eta})$ with
$(\hat{\eta}_q\hat{\eta}_p + \hat{\eta}_p\hat{\eta}_q= {1\over 2}
\delta_{pq}$) satisfying differential  equations enforcing the
quantum counterparts of the constraints (\ref{SO(9)=}).

{\bf Acknowledgments}. Support from the Ministerio de Educaci\'on
y Ciencia and EU FEDER funds (FIS2005-02761), the Generalitat
Valenciana, INTAS (2006-7928), the EU (MRTN-CT-2004-005104) and
the Ukrainian State Fund for Fundamental Research (383) is
gratefully acknowledged.


{\small

\begin{thebibliography}{99}

 \bibitem{NBloops}
N.~Berkovits, {\it Covariant quantization of the supermembrane},
  JHEP {\bf 0209}, 051 (2002)
  [hep-th/0201151];
  {\it Multiloop amplitudes and vanishing theorems using the pure spinor
  formalism for the superstring},
  JHEP {\bf 0409}, 047 (2004) [hep-th/0406055];
  {\it Super-Poincare covariant two-loop superstring amplitudes},
  JHEP {\bf 0601}, 005 (2006)
  [hep-th/0503197];
  {\it New higher-derivative $R^4$ theorems},
  [hep-th/0609006].

\bibitem{pure-GS}
I.~Oda and M.~Tonin,
  {\it On the Berkovits covariant quantization of GS superstring},
  Phys.\ Lett.\ {\bf B520}, 398 (2001)
  [hep-th/0109051];  \\
  N.~Berkovits and D.~Z.~Marchioro,
  {\it Relating the Green-Schwarz and pure spinor formalisms for the
  superstring},
  JHEP {\bf 0501}, 018 (2005)
 [hep-th/0412198].

\bibitem{Dima+Mario+02}
  M.~Matone, L.~Mazzucato, I.~Oda, D.~Sorokin and M.~Tonin,
  {\it The superembedding origin of the Berkovits pure spinor covariant
  quantization of superstrings},
  Nucl.\ Phys.\ B {\bf 639}, 182 (2002)
  [hep-th/0206104].

\bibitem{bpstv}
  I.~A.~Bandos, D.~P.~Sorokin, M.~Tonin, P.~Pasti and D.~V.~Volkov,
  {\it Superstrings and supermembranes in the doubly supersymmetric geometrical
  approach},
  Nucl.\ Phys.{\bf B446}, 79 (1995)
  [hep-th/9501113].


\bibitem{Dima}
D.~P.~Sorokin,
  {\it Superbranes and superembeddings},
  Phys.\ Rept.\  {\bf 329}, 1 (2000) [hep-th/9906142] and refs. therein.


\bibitem{nonmNB}
P.~A.~Grassi, G.~Policastro, M.~Porrati and P.~Van Nieuwenhuizen,
{\it Covariant quantization of superstrings without pure spinor
constraints},
  JHEP {\bf 0210}, 054 (2002)
  [hep-th/0112162]; \\
I.~Oda and M.~Tonin, {\it On the b-antighost in the pure spinor
quantization of superstrings},
  Phys.\ Lett.\ B {\bf 606}, 218 (2005) [hep-th/0409052]; \\
 N.~Berkovits and C.~R.~Mafra,
 {\it  Some superstring amplitude computations with the non-minimal pure spinor
  formalism}, JHEP {\bf 0611}, 079 (2006) [hep-th/0607187].

\bibitem{Sok}
E.\ Sokatchev, {\it Light-cone harmonic superspace and its
applications}, Phys. Lett. {\bf B169}, 209 (1987);  {\it Harmonic
superparticle}, Class. Quantum Grav. {\bf 4}, 237-246 (1987).

\bibitem{Lharm}
E. Nissimov, S. Pacheva, S. Solomon, {\it Covariant first and
second quantization of the N = 2, D = 10 Brink-Schwarz
superparticle},
  {Nucl. Phys.} {\bf B296}, 462-492 (1988) ;
 {\it Off-Shell Superspace D = 10 Superyang-Mills From Covariantly
 Quantized Green-Schwarz Superstring}, {\it ibid.} {\bf B317}, 344 (1989);\\
 R.  Kallosh and M. Rahmanov,
 {\it Covariant quantization of the Green-Schwarz superstring},
 {Phys. Lett.} {\bf B209}, 233-238 (1988);
 {\it Consistency of covariant quantization of the GS-string},
 {\it ibid.}
 {\bf B214}, 549-554 (1988);\\
P. Wiegmann, {\it Extrinsic geometry of superstrings},
 Nucl. Phys. {\bf  B323}, 330-336 (1989);\\
 I.~A.~Bandos, {\it A superparticle in Lorentz-harmonic superspace},
  Sov. J.\ Nucl. Phys.  {\bf 51}, 906 (1990);
{\it   Multivalued action functionals, Lorentz harmonics, and
spin},
  JETP Lett. {\bf 52}, 205 (1990).

\bibitem{Ghsds}
  A.~Galperin, P.S. Howe and K.S. Stelle,  {\it The superparticle and the Lorentz group},
  Nucl. Phys. {\bf B368},  248-280  (1992);\\
   F. Delduc, A.~Galperin and E. Sokatchev,
   {\it Lorentz-harmonic (super)fields and (super)particles},
   Nucl. Phys. {\bf B368}, 143-171 (1992).

\bibitem{BZ}
I.~A.~Bandos and A.~A.~Zheltukhin, {\it Green-Schwarz Superstrings
In Spinor Moving Frame Formalism},
  Phys.\ Lett.\ {\bf B288}, 77 (1992);\\
  {\it N=1 superp-branes in twistor-like Lorentz harmonic
  formulation},
  Class.\ Quant.\ Grav.\  {\bf 12}, 609 (1995) [hep-th/9405113].

\bibitem{BZstrH}
I.~A.~Bandos and A.~A.~Zheltukhin, {\it D = 10 Superstring:
lagrangian and hamiltonian mechanics in twistor - like Lorentz
harmonic formulation},
  Phys. Part. Nucl.  {\bf 25}, 453 (1994).

\bibitem{GHT93} A.~S.~Galperin, P.~S.~Howe and P.~K.~Townsend,
{\it Twistor transform for superfields},
 Nucl. Phys. {\bf B402}, 531-547 (1993).

\bibitem{JdA00b}
I.~A.~Bandos, J.~A.~de Azc\'arraga, M.~Pic\'on and O.~Varela,
 {\it  D = 11 superstring model with 30 kappa-symmetries and 30/32 BPS states  in
  an extended superspace},
  Phys.\ Rev. {\bf D69}, 085007 (2004)
 [hep-th/0307106].

\bibitem{Nicolai87}
H.~Nicolai,
 {\it D = 11 Supergravity With Local SO(16) Invariance},
  Phys.\ Lett.\ B {\bf 187}, 316 (1987);
  {\it On M-theory},
  hep-th/9801090.

\bibitem{B+T=D-branes}
E.~Bergshoeff and P.~K.~Townsend, {\it Super D-branes},
  Nucl. Phys.  {\bf B490}, 145 (1997) [hep-th/9611173].

\bibitem{BL98'} I.~A.~Bandos and J.~Lukierski,
{\it  New superparticle models outside the HLS supersymmetry
scheme},
  Lect. Not. Phys. {\bf 539}, 195 (2000)
  [hep-th/9812074].

\bibitem{van2}
J.W. van Holten and A. van Proeyen, {\it N=1 Supersymmetry
algebras in $D = 2$, $D = 3$, $D = 4$ mod $8$}, J. Phys. {\bf
A15}, 3763 (1982).

\bibitem{BL98}
I. Bandos and J. Lukierski, {\it Tensorial central charges and new
superparticle models with fundamental spinor coordinates}, Mod.
Phys. Lett. {\bf 14}, 1257-1272 (1999) [hep-th/9811022].

\bibitem{JdA00}
C. Chryssomalakos, J.A. de Azc\'{a}rraga, J.M. Izquierdo and J.C.
P\'{e}rez Bueno, {\it The geometry of branes and extended
superspaces}, Nucl. Phys. {\bf B567}, 293-330 (2000)
[hep-th/9904137].

\bibitem{BLS99+BBdAST04}%
I. Bandos, J. Lukierski and D. Sorokin, {\it Superparticle Models
with Tensorial Central Charges}, Phys. Rev. {\bf D61},  045002
(2000) [hep-th/9904109];
\\
I.A.~Bandos, X. Bekaert, J.~A.~de Azc\'arraga, D. Sorokin and M.
Tsulaia,
{\it Dynamics of higher spin fields and tensorial space},
JHEP {\bf 0505}, 031 (2005) [hep-th/0501113].

\bibitem{BPS01}
 I.A. Bandos, J.A. de Azc\'arraga, J.M. Izquierdo and J. Lukierski,
{\it BPS states in M-theory and twistorial constituents},
Phys. Rev. Lett. {\bf 86}, 4451 (2001) [hep-th/0101113]; \\
I.~A.~Bandos, J.~A.~de Azc\'arraga, J.~M.~Izquierdo, M.~Pic\'on
and O.~Varela,{\it  On BPS preons, generalized holonomies and D =
11 supergravities},
  Phys.\ Rev. {\bf D69}, 105010 (2004)
  [hep-th/0312266].

\bibitem{gs92}
  I.~A.~Bandos, M.~Cederwall, D.~P.~Sorokin and D.~V.~Volkov,
{\it Towards a complete twistorization of the heterotic string},
Mod. Phys. Lett. {\bf A9}, 2987 (1994) [hep-th/9403181] and refs.
therein.

\bibitem{Green+99}
  M.~B.~Green, M.~Gutperle and H.~H.~Kwon,
{\it Light-cone quantum mechanics of the eleven-dimensional
superparticle},
  JHEP {\bf 9908}, 012 (1999) [hep-th/9907155].

 \bibitem{F78S83}
A. Ferber, {\it Supertwistors and conformal supersymmetry},
Nucl.\ Phys. {\bf B132}, 55-64 (1978); \\
 T. Shirafuji,
{\it Lagrangian mechanics of massless particles with spin}, Prog.
Theor. Phys. {\bf 70}, 18 (1983).

\bibitem{Dima88}
 D.~P.~Sorokin, {\it Supersymmetric particles,
classical dynamics and its quantization}, preprint ITP-87-159,
Kiev, 1988 [unpublished]; for a discussion see
\cite{BLS99+BBdAST04}.

\bibitem{Itzhak+MP05+IB+JdA+C06}
I.~Bars and M.~Pic\'on, {\it Twistor transform in d dimensions and
a unifying role for twistors},
  Phys.\ Rev. {\bf D73}, 064033 (2006)
  [hep-th/0512348];\\
I.~A.~Bandos, J.~A.~de Azc\'arraga and C.~Miquel-Espanya, {\it
Superspace formulations of the (super)twistor string},
  JHEP {\bf 0607}, 005 (2006)
  [arXiv:hep-th/0604037].



\end{thebibliography}
}
\end{document}